\documentclass[aps,prl,twocolumn,groupedaddress,floatfix,showpacs]{revtex4}

\usepackage{graphicx}
\usepackage{amsmath}
\usepackage{psfrag}
\begin{document}
\newcommand{\al}{\alpha}
\newcommand{\gam}{\gamma}
\newcommand{\eps}{\epsilon}
\newcommand{\lam}{\lambda}
\newcommand{\om }{\omega}
\newcommand{\omr}{\om_r}

\ProvideTextCommandDefault{\textonehalf}{${}^1\!/\!{}_2\ $}

\title{Dressed Spin of Helium-3}

\author{A. Esler}
\email[Electronic address: ]{asharp@uiuc.edu}
\author{J. C. \surname{Peng}}
\author{D. Chandler}
\author{D. Howell}
\affiliation{Department of Physics, University of Illinois
  at Urbana-Champaign, Urbana, Illinois 61801}

\author{S. K. Lamoreaux} 
\altaffiliation[Present address: ]{Yale University,
  New Haven, CT 06520.}
\author{C. Y. Liu}
\altaffiliation[Present address: ]{Indiana University,
  Bloomington, IN 47405.}
\author{J. R. Torgerson}
\affiliation{Physics Division, Los Alamos National
  Laboratory, Los Alamos, New Mexico 87545}

\date{\today}

\begin{abstract}
  We report a measurement of dressed-spin effects of polarized
  ${}^{3}$He atoms from a cold atomic source traversing a region of
  constant magnetic field $B_0$ and a transverse oscillatory dressing
  field $B_d\cos \omega_d t$.  The observed effects are compared with
  a numerical simulation using the Bloch equation as well as a
  calculation based on the dressed-atom formalism. An application of
  the dressed spin of ${}^{3}$He for a proposed neutron electric
  dipole moment measurement is also discussed.
\end{abstract}



\maketitle

The existence of a permanent electric dipole moment (EDM) of an
elementary particle such as the neutron is direct evidence for
time-reversal (T) symmetry breaking, which implies a violation of CP
symmetry assuming CPT invariance~\cite{khriplovich97}.  Although CP
violation is known to occur in neutral K and B meson systems, it has
never been found for hadrons consisting of light quarks only, such as
the neutron.  Therefore, observation of a nonzero neutron EDM would
provide qualitatively new information on the origin of CP violation.

The current experimental upper limit of the neutron EDM ($d_n$),
obtained from an experiment~\cite{baker06} using bottled ultracold
neutrons (UCNs), is $|d_n|<2.9 \times 10^{-26}e$-cm (90\% C.L.); see
also~\cite{BAKER06comment}.  A new experimental search for the neutron
EDM has been proposed using UCNs produced and trapped in a bath of
superfluid ${}^{4}$He~\cite{GOLUB94,EDMPROPOSAL}.  The experiment
searches for a shift of the UCN precession frequency due to the
interaction of $d_n$ with an applied electric field.

In the proposed neutron EDM experiment, a small concentration of
polarized ${}^{3}$He atoms ($X \sim 10^{-10}$) would be introduced
into the superfluid to serve as a comagnetometer. The ${}^{3}$He atoms
would also function as a highly sensitive spin analyzer due to the
large difference between the $n$-${}^{3}$He absorption with total spin
$J=0$ compared to $J=1$~\cite{PASSELL66}.  The absorption reaction $n\
+ {}^{3}\text{He}\to p\ +{}^{3}\text{H}\ $ releases 764~keV of total
kinetic energy.  This recoil energy excites short-lived molecules in
the superfluid ${}^{4}$He which emit ultraviolet scintillation
light~\cite{SURKO1970}.  Consequently, the observed rate of
scintillations depends on the relative angle between the UCN and
${}^{3}$He spins.  In a transverse magnetic field $B_0$, the UCN and
${}^{3}$He spins will precess at their respective Larmor frequencies:
$\om_n = \gam_n B_0$, and $\om_3 = \gam_3 B_0$ where $\gam_i$ is the
gyromagnetic ratio of each species. If the ${}^{3}$He and UCN spins
are parallel at time $t=0$, a relative angle between the spins
develops over time because the ${}^{3}$He magnetic moment is larger
than that of the neutron ($\gam_3 \approx 1.1\, \gam_n$).  Therefore
the rate of scintillations observed is modulated at the difference of
the two spin precession frequencies:
\begin{equation} \om_\text{rel} =
  (\gam_3 - \gam_n)B_0 \approx 0.1\, \gam_n  B_0.
  \label{eq:omegarel}
\end{equation}
In the presence of a static electric field $E$ parallel to $B_0$,
Eq.~\ref{eq:omegarel} gains a term proportional to the neutron EDM:
\begin{equation} \om_\text{rel} = (\gam_3 - \gam_n)B_0 
  +\, 2 d_n E/\hbar.
  \label{eq:omegarelwithE}
\end{equation}
Eq.~\ref{eq:omegarelwithE} shows that $\om_\text{rel}$ depends only on
$d_n E$ in the limit of $B_0 \to 0$. Alternatively, the experimental
signal would become independent of $B_0$ if the condition $\gam_3 -
\gam_n = 0$ were satisfied. Spurious signals due to inhomogeneity or
slow drifts in the magnetic fields would thereby be eliminated. The
UCN and ${}^{3}$He magnetic moments can be modified, and in fact
equalized, by the dressed spin effect~\cite{GOLUB94, CARGESE,
  HAROCHE70b} in which a particle's effective magnetic moment is
modified by applying an oscillating magnetic field $B_d \cos \om_d t$
perpendicular to $B_0$. In the weak-field limit ($B_0 \ll \om_d/\gam$,
or $y \ll 1$ where $y\equiv \gam B_0 / \om_d$), Polonsky and
Cohen-Tannoudji~\cite{POLONSKY65} showed that the dressed magnetic
moment $\gam_i'$ is given by
\begin{equation}\gam_i' = \gam_i J_0(x_i), \qquad x_i \equiv \gamma_i
  B_d/\om_d,
\label{eq:J0relation}
\end{equation}
where $J_0$ is the zeroth-order Bessel function of the first kind, and
$x_i$ is a dimensionless parameter proportional to the dressing field
strength. Using this expression, one can solve for the ``critical''
dressing field magnitude which makes $\gam_n' = \gam_3'$. If this
critical dressing field is applied, corresponding to $x_3 =
1.32$~\cite{GOLUB94}, the relative precession between the UCN and
${}^{3}$He (Eq.~\ref{eq:omegarelwithE}) vanishes except for the
contribution from $d_n E$. In addition, modulating the $x$ parameter
with a different frequency $\om_\text{m}$ causes the observed
scintillation rate to have a first harmonic term with amplitude
proportional to the neutron EDM and the applied electric
field~\cite{GOLUB94}.

\begin{figure*}
  \psfrag{spin}{\!\!\!\!$\pi/2$ spin rotation coil}
  \psfrag{pol}{spin polarizer}
  \psfrag{ana}{spin analyzer}
  \psfrag{rga}{RGA}
  \psfrag{gf}{guide fields}
  \psfrag{gf2}{guide fields}
  \psfrag{Ls}{$L_s = 90$ cm solenoid}
  \psfrag{Lr}{$\ell_r = 45$ cm}
  \psfrag{Ld}{$\ell_d = 8.6$ cm}
  \psfrag{beam}{\!\!\!\!\!\!${}^{3}$He beam}
  \psfrag{dress}{dressing coils}
  \includegraphics[width=0.8\textwidth]{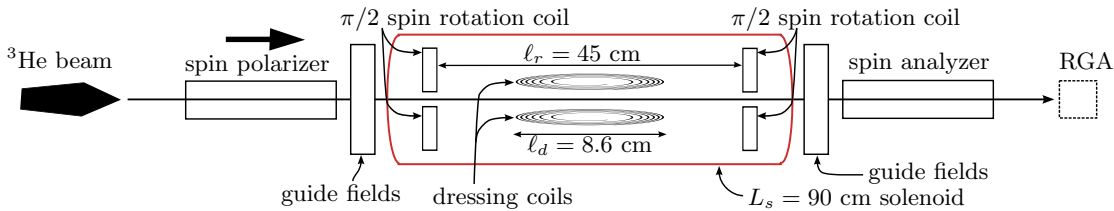}
  \caption{\label{fig:setup}(Color online) Schematics of the apparatus
    used for measuring the ${}^{3}$He resonance frequency. In this
    coordinate system, the beam propagates along $\mathbf{\hat{z}}$
    and $\mathbf{\hat{x}}$ points vertically upwards.}
\end{figure*}

Modification of the neutron magnetic moment using an oscillatory
magnetic field has been demonstrated experimentally by Muskat,
Dubbers, and Sch\"{a}rpf~\cite{MUSKAT87}. Other authors have described
work using excited states of mercury or alkali
atoms~\cite{experiments}, but the effects of rf spin dressing on
${}^{3}$He nuclei have never been reported.  In this paper we present
results of an experiment on polarized ${}^{3}$He which demonstrate
changes of the ${}^{3}$He dressed magnetic moment as predicted by
Eq.~\ref{eq:J0relation} for small values of $y$.  Deviations from
Eq.~\ref{eq:J0relation} are observed for larger values of $y$.
Numerical calculations using both classical and quantum mechanical
methods are compared with the experimental results.

We measured the dressed ${}^{3}$He precession frequency using the Ramsey
separated oscillatory fields (SOF) method~\cite{RAMSEY56} and the
experimental apparatus shown in Fig.~\ref{fig:setup}.  Cold ${}^{3}$He
atoms from an effusive beam source at $\sim$1.0 K were polarized by a
strong (7.5~kG) quadrupole magnetic field and entered a 90-cm long
solenoid with 99.5\% polarization~\cite{LAMVELOCITYDIST} along the
direction of the solenoid field $B_0\, \mathbf{\hat{z}}$. Two pairs of
$\pi/2$ coils were placed inside the solenoid to provide the
vertical fields $B_r \cos\omr t\, \mathbf{\hat{x}}$
required for the SOF method. In addition, an independent pair of coils
was located at the middle of the solenoid to provide the vertical
dressing field. Downstream of the solenoid, a spin analyser identical
to the quadrupole polarizer transmitted those ${}^{3}$He atoms which
were polarized along the $\mathbf{\hat{z}}$ direction. A residual gas
analyser (RGA) then counted the flux of the transmitted ${}^{3}$He
atoms.

The ${}^{3}$He beam from the source had a thermalized velocity
distribution $f(v)$ (a Maxwellian distribution modified by the
polarizer's acceptance) which was determined from a measurement of the
${}^{3}$He beam transmission with a single rf coil set at the Larmor
frequency. We found that $f(v)$ peaked at $\overline{v}$ =
155~m~s$^{-1}$ with FWHM = 70~m~s$^{-1}$.

\begin{figure}
  \psfrag{xlabel}{$\pi/2$ coil frequency (kHz)}
  \psfrag{ylabel}{Transmitted ${}^{3}$He flux (arbitrary unit)}
  \includegraphics[width=7cm]{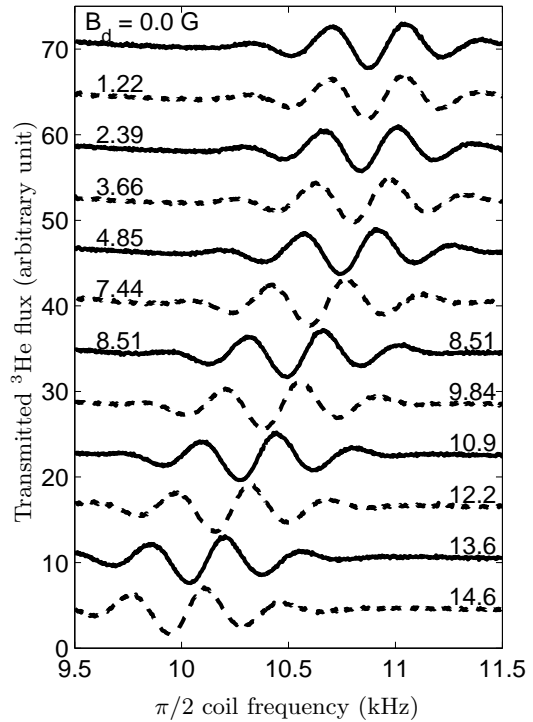}
  \caption{\label{fig:wiggleplot}${}^{3}$He beam transmission vs.\ SOF
    frequency showing shift of the resonance frequency with increasing
    magnitude of the dressing field $B_d$.  Sequential data traces are
    offset vertically for clarity.  In these data $B_0 = 3.36$~G,
    $\omega_0/2\pi = 10.89$~kHz, and $\omega_d/2\pi = 29.5$~kHz ($y =
    0.369$).}
\end{figure}

\begin{figure}
  \begin{center}
    \psfrag{xlabel}{{$\quad$ Dressing field $B_d$ (G)}}
    \psfrag{ylabel}
    {$\quad$ Resonance frequency $\om'_r/2\pi$ (kHz)}
    \psfrag{omhi}{\large{$B_0$ = 8.50 G}}
    \psfrag{omlo}{\large{$B_0$ = 3.36 G}}
    \includegraphics[width=0.45\textwidth]{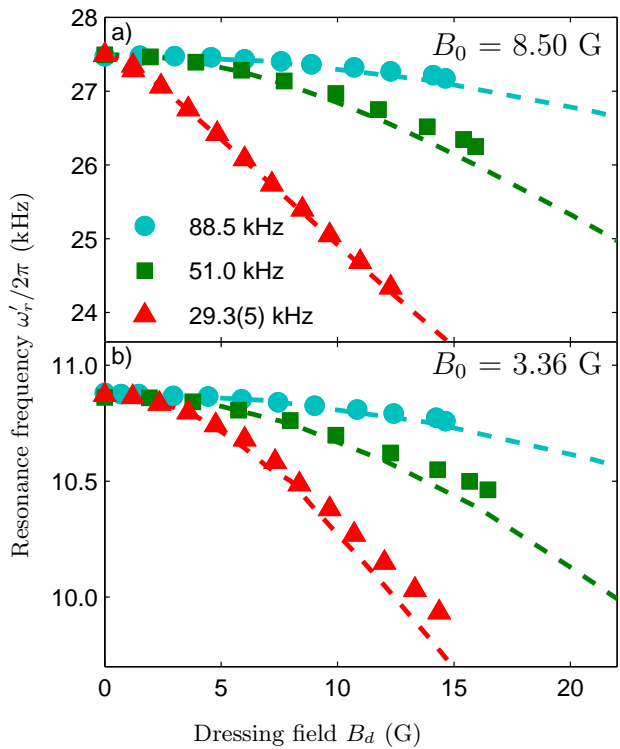}
   \end{center}
   \caption{\label{fig:dressingB}(Color online) ${}^{3}$He resonance
     frequency data as a function of $B_d$ for several dressing field
     frequencies $\om_d$. (a) $B_0$ = 8.50~G.  (b) $B_0$ = 3.36~G.
     The triangle symbol indicates $\om_d/2\pi$ = 29.3~kHz in (a) and
     29.5~kHz in (b).  Dashed lines show the results of Bloch
     simulations for each frequency.}
\end{figure}

In the dressed spin measurement, the frequency of the rf fields
$(\,\omr)$ was varied near the ${}^{3}$He Larmor precession
frequency, producing oscillations in the transmitted atom flux as
shown in Fig.~\ref{fig:wiggleplot} (a detailed discussion of the SOF
method is given in~\cite{RAMSEY56}).  With the dressing field off, the
global minimum in the transmitted flux is observed at the ${}^{3}$He's
Larmor frequency (this is the ordinary resonance condition
$\omr$ = $\om_0$ = $\gam B_0$). The shape of the transmission
curve is consistent with the velocity-averaged transition probability
calculated using the measured atomic velocity distribution $f(v)$.
When the dressing field $B_d \cos \omega_d t\, \mathbf{\hat{x}}$ was
applied, the value of $\omr$ that produced the minimum RGA flux
shifted to a different frequency, as demonstrated in
Fig.~\ref{fig:wiggleplot}.  Measurements were performed at two $B_0$
values (3.36~G and 8.50~G) and the dressing field frequency and
magnitude were varied as parameters. Several different dressing field
frequencies were investigated for each $B_0$ setting, and at each
frequency the dressing field's magnitude was varied over 10 to 15
values ranging from 0 up to 15-20~G.  For each $B_0$, $B_d$, and
$\om_d$ combination the frequency of the $\pi/2$ coils at which the
RGA flux reached its minimum was determined. The results of our
measurements are plotted in Fig.~\ref{fig:dressingB}.

The shifts observed in the ${}^{3}$He resonance frequency are due to
changes in the effective magnetic moment of the ${}^{3}$He caused by
the dressing field. In the undressed, or ``free'' case ($B_d=0$) the
transverse components of the ${}^{3}$He spin precess about $B_0$ at
the Larmor frequency $\om_0=\gam B_0$ during the transit time between
the two $\pi/2$ rotation coils, which we call $t_r$. The resonance
condition occurs when the spin precesses in phase with the rf field,
i.e.~$\gam B_0 t_r = \omr t_r$. Modifying the ${}^{3}$He
magnetic moment with the dressing field caused the spin to precess
with a different frequency in the region of the dressing field, and
therefore the observed magnetic resonance occurred at a frequency
$\om'_r$ different from the undressed case.  Writing $\gam'$ for
the gyromagnetic ratio of the dressed ${}^{3}$He, the total spin
precession of the ${}^{3}$He is a sum of two contributions -- the
precession inside the dressing field with frequency $\gam' B_0$, and
the free precession with frequency $\om_0 = \gam B_0$ outside the
dressing region.  The resonance occurs when the phase angle of the rf
field is equal to the total ${}^{3}$He precession angle:
\begin{equation}
  \om'_r t_r = \gam B_0 (t_r-t_d)+\gam' B_0\, t_d
\end{equation}
and the resonance frequency shift $\Delta \omr =
\omr'-\om_0$\, is
\begin{equation}
  \Delta \omr = B_0(\gam'-\gam)\frac{t_d}{t_r} = \,
  \om_0 (\gam'/\gam-1)\frac{\ell_d}{\ell_r} \label{eq:deltaomegaR}
\end{equation}
where $t_d$ is the ${}^{3}$He's time of flight through the dressing
region of length $\ell_d$, and $\ell_r$ is the separation of the two
$\pi/2$~fields. Secondary minima in the transmission signal are
observed (Fig.~\ref{fig:wiggleplot}) when the precession and rf phase
difference is an integer multiple of 2$\pi$.

We have used both classical and quantum-mechanical approaches to
interpret the experimental results. In the first case, we numerically
integrate the Bloch equation with a fourth-order Runge-Kutta method
to propagate the ${}^{3}$He through the solenoid region and determine
the final polarization $\mathbf{s} \cdot \mathbf{\hat{z}}$. We
simulated all of the measurements by varying the SOF frequency about
the ${}^{3}$He's Larmor frequency and averaging the result over
$f(v)$. The resonance curves thus obtained are in good agreement with
our experimental data and in particular, the resonance frequency
shifts due to the dressing field are well reproduced, as shown in
Fig.~\ref{fig:dressingB}.

In addition to the classical simulations, we have also interpreted the
experimental observations using the dressed atom approach pioneered by
Cohen-Tannoudji \textit{et al.}~\cite{CARGESE}.  The Hamiltonian of a
spin-\textonehalf particle with gyromagnetic ratio $\gam$ subjected to
the constant magnetic field $B_0\, \mathbf{\hat{z}}$ and a linearly
polarized rf field $B_d \cos \om_d t\, \mathbf{\hat{x}}$ can be
written
\begin{equation} \hat{H} = -\gam B_0 \hat{S}_z + \hbar
  \om_d\, \hat{a}^{\dagger} \hat{a} + \lam \hat{S}_x(\hat{a} +
  \hat{a}^{\dagger}),
  \label{eq:hamil}
\end{equation}
where $\hat{S}_x$ and $\hat{S}_z$ are the spin operators along
$\mathbf{\hat{x}}$ and $\mathbf{\hat{z}}$ respectively ($\hat{S}_z$
having eigenvalues $m_z = \pm \frac{1}{2} \hbar$).  The first term of
Eq.~\ref{eq:hamil} is the Zeeman interaction, and the second term is
the energy of the dressing field with creation and annihilation
operators $\hat{a}^{\dagger}$ and $\hat{a}$.  The final term in
Eq.~\ref{eq:hamil} describes the coupling of the particle's spin to
the photon field with strength $\lambda = \gamma B_d
/2\sqrt{\overline{n}}$, where $\overline{n} \gg 1$ is the average
number of photons.  This interaction term allows the particle to absorb or
emit photons which entails the exchange of energy and angular
momentum.  Because the rf field is perpendicular to $B_0$ (and can be
decomposed into a superposition of right- and left-handed circularly
polarized fields), only $\Delta m_z = \pm \hbar$ transitions are allowed.

\begin{figure}[t]
  \psfrag{ylabel}{ {Resonance frequency shift $\Delta \omr/2\pi$ (kHz)}}
  \psfrag{xlabel}{ {Dressing parameter $x$}}
  \psfrag{omhi}{\large{$B_0$ = 8.50 G}}
  \psfrag{omlo}{\large{$B_0$ = 3.36 G}}
  \includegraphics[width=0.45\textwidth]{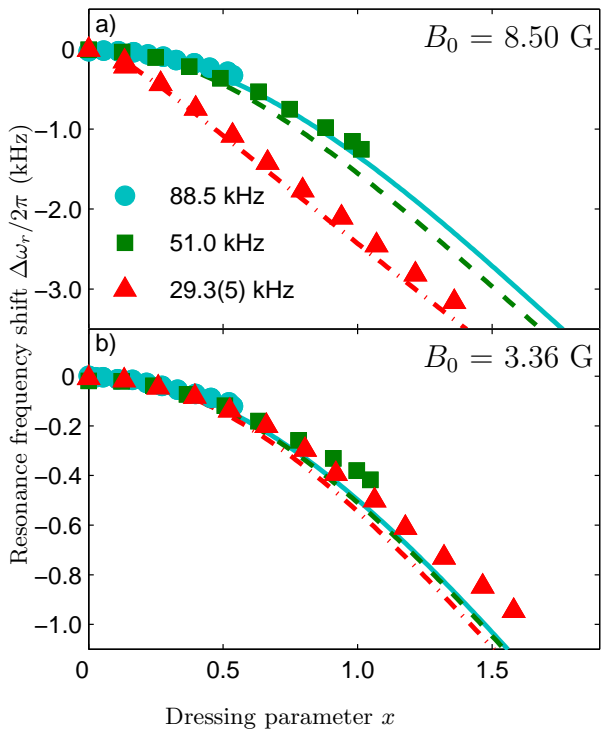}
  \caption{\label{fig:dressingX}(Color online) Change of the
    ${}^{3}$He precession frequency as a function of the dressing
    parameter $x = \gamma B_d / \om_d$.  The curves show the expected
    resonance frequency shifts computed from the dressed spin energy
    spectrum and Eq.~\ref{eq:deltaomegaR}. The values of $y$ (from top
    to bottom) in (a) are 0.31, 0.54, and 0.94; in (b), the values are
    0.12, 0.21, and 0.37.}  %
\end{figure}

In the weak-field regime ($B_0 \ll \om_d/\gam$, or $y \ll 1$),
Eq.~\ref{eq:hamil} can be solved analytically with
the result $\gam'=\gam J_0(x)$~\cite{POLONSKY65}.
Eq.~\ref{eq:deltaomegaR} implies that the resonance frequency shift
then becomes
\begin{equation}
  \Delta \omr = \om_0\left[J_0(x)-1\right] \frac{\ell_d}{\ell_r},
\label{eq:deltaomegaRsimple}
\end{equation}
which only depends on the dressing strength $x = \gam B_d / \om_d$.
The experimental values of $\Delta \omr$ are plotted as a
function of $x$ in Fig.~\ref{fig:dressingX} for measurements at two
different $B_0$ settings and several values of $y$. As shown in
Fig.~\ref{fig:dressingX}(b), this ``$x$-scaling'' behavior is indeed
observed for $\Delta \omega_r$ measured at $B_0$ = 3.36~G where
the $y$ values are small. For data taken at $B_0$ = 8.50~G, deviation
from $x$-scaling is clearly observed for the measurement with
$\om_d/2\pi$ = 29.3~kHz ($y=0.94$), for which the expression $\gam' =
\gam J_0(x)$ no longer holds.

\begin{figure}
  \psfrag{Dressed spin energy}{Dressed spin energy $E/\hbar\om_d$}
  \psfrag{ylabel}{{\!\!\!\!\!\!Dressed spin energy $E/\hbar\om_d -
      \overline{n}$}}
  \psfrag{xlabel}{{$y = \gamma B_0/\omega_d$}}
  \psfrag{dE}{  $\Delta E$}
  \includegraphics[width=0.43\textwidth]{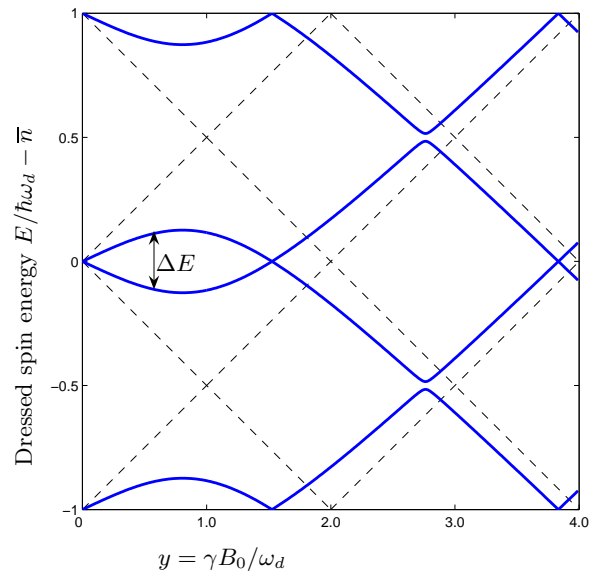}
  \caption{\label{fig:Ediag}(Color online) Sample energy diagram of
    the dressed spin system calculated as a function of $y$, for
    dressing parameter $x=1.57$.  Dashed lines indicate the Zeeman
    splittings in the undressed system $(E_0 = \pm \frac{1}{2} \hbar
    \om_0)$. The energy scale is given in units of the dressing field
    photon energy $\hbar \om_d$.}
\end{figure}

To understand the observed deviation from $x$-scaling shown in
Fig.~\ref{fig:dressingX}, we have calculated the dressed spin energy
diagram by diagonalizing the full Hamiltonian of Eq.~\ref{eq:hamil}.
In Fig.~\ref{fig:Ediag} we show an example of the dressed energy
eigenvalues $E$ as a function of the static field $B_0$, for a
constant dressing field magnitude corresponding to $x = 1.57$, which
is the largest value acheived in our measurements.  The diagram shows
how the Zeeman splitting in the undressed system is modified by the
presence of the dressing field~\cite{CARGESE}.  From the energy
difference $\Delta E$ between the dressed eigenstates, $\gam'$ is
given by $\Delta E/B_0$ and $\Delta \omr$ is calculated using
Eq.~\ref{eq:deltaomegaR} to obtain the curves in
Fig.~\ref{fig:dressingX}. These results show that the observed
deviation from $x$-scaling can be quantitatively described in this
quantum mechanical approach.

In summary, we have measured the modification of the precession
frequency of a polarized ${}^{3}$He beam in a constant magnetic field
superimposed by a transverse oscillating dressing field. In the
weak-field limit ($y = \gam B_0/\om_d \ll 1$), the modified
gyromagnetic ratio $\gam'$ obeys the relation $\gam' = \gam J_0(x)$.
Deviation from this relation is observed at larger values of $y$. The
observed modification of the ${}^{3}$He effective gyromagnetic ratio
can be well described by classical calculations using the Bloch
equation as well as by the quantum approach based on the dressed-atom
formalism.  This result supports the proposal to use a dressing field
to modify the neutron and ${}^{3}$He precession frequencies in a
neutron EDM experiment.

This work was supported in part by the U.S. Department of Energy
and the National Science Foundation.

\end{document}